\documentclass[doublecol]{epl2}

\title{On the Explanation of the Paramagnetic Meissner Effect in Superconductor/Ferromagnet Heterostructures}

\author{B. Nagy\inst{1,2}  \and Yu. Khaydukov\inst{3,4,5}  \and D. Efremov\inst{3,6} \and A. S. Vasenko\inst{7} \and L. Mustafa\inst{3} \and J.-H. Kim\inst{3} \and T. Keller\inst{3,4} \and K. Zhernenkov\inst{8} \and A. Devishvili\inst{8,9} \and R. Steitz\inst{10} \and B. Keimer\inst{3} \and L. Botty\'{a}n\inst{4}}
\shortauthor{B. Nagy \etal}

\institute{
  \inst{1} Wigner Research Centre for Physics, Hungarian Academy of Sciences, Budapest, Hungary\\
  \inst{2} Link\"{o}ping University, Link\"{o}ping, Sweden\\
  \inst{3} Max-Planck Institute for Solid State Research, Stuttgart, Germany\\
  \inst{4} Max Planck Society Outstation at FRM-II, Garching, Germany\\
  \inst{5} Skobeltsyn Institute of Nuclear Physics, Moscow State University, Moscow, Russia\\
  \inst{6} Leibniz Institute for Solid State and Materials Research Dresden, Germany\\
  \inst{7} National Research University Higher School of Economics,  Moscow, Russia\\
  \inst{8} Ruhr-Universit\"{a}t Bochum, Bochum, Germany\\
  \inst{9} Division of Physical Chemistry, Department of Chemistry, Lund University,  Lund, Sweden\\
  \inst{10} Helmholtz-Zentrum f\"{u}r Materialien und Energie, Berlin, Germany
}

\abstract{
An increase of the magnetic moment in superconductor/ferromagnet (S/F) bilayers V(40nm)/F [F$=$Fe(1,3nm), Co(3nm), Ni(3nm)] was observed using SQUID magnetometry upon cooling below the superconducting transition temperature $T_C$ in magnetic fields of 10 Oe to 50 Oe applied parallel to the sample surface. A similar increase, often called the paramagnetic Meissner effect (PME), was observed before in various superconductors and superconductor/ferromagnet systems. To explain the PME effect in the presented S/F bilayers a model based on a row of vortices located at the S/F interface is proposed. According to the model the magnetic moment induced below $T_C$ consists of the paramagnetic contribution of the vortex cores and the diamagnetic contribution of the vortex-free region of the S layer. Since the thickness of the S layer is found to be 3-4 times less than the magnetic field penetration depth, this latter diamagnetic contribution is negligible. The model correctly accounts for the sign, the approximate magnitude and the field dependence of the paramagnetic and the Meissner contributions of the induced magnetic moment upon passing the superconducting transition of a ferromagnet/superconductor bilayer.}

\begin{document}

\maketitle

The paramagnetic Meissner effect (PME), i.e. the appearance of a positive (rather than a negative) moment below the superconducting transition temperature $T_C$ was first observed in high-$T_C$ superconductors (HTSC) cooled in small fields (less than $\sim$ 5 Oe) and initially was explained by peculiarities of the d-wave electron coupling in those systems. Later, however, PME was observed in conventional superconductors, like Al or Nb, which required a less exotic explanation (see review \cite{Li} on the PME in bulk superconductors). One of such explanations is based on flux trapping below $T_C$ with further compression upon cooling \cite{Koshelev, Moshchalkov}.

A similar effect, the increase of magnetic moment, was also observed recently in superconductor/ ferromagnet heterostructures (S/F) of different compositions\cite{Satapathy, Bernardo, Torre, Chen, Wu, Monton, Ovsyannikov} with both HTSC and conventional superconductors. For the explanation of the PME in S/F structures different models based on electrodynamical or exchange coupling mechanisms were proposed. In the former case the presence of the stray field produced by the F layer and the response of the superconductor to this field is usually considered \cite{Chen, Wu, Monton}. The second model requires the presence of exchange coupled electrons on the S/F interface \cite{Satapathy, Bernardo, Ovsyannikov} due to the proximity effect. At the moment there is no unanimous opinion regarding the nature of the PME in S/F structures.

In our previous studies an increase of the magnetic moment was observed in a S/F bilayer consisting of V(40nm)/Fe(1nm) layers measured with SQUID and polarized neutron reflectometry (PNR) \cite{Khaydukov1, Khaydukov2}. Comprehensive analysis of the neutron and SQUID data showed the appearance of an additional positive magnetic moment on the S layer side of the S/F interface. The 7 nm thickness of this magnetic sublayer is comparable with the superconducting coherence length $\xi_S = 9$ nm derived from transport measurements. In order to explore the possible dependence of the PME on the magnetic exchange field and the layer thickness, a series of samples were subsequently prepared with different F layers in contact with the same S layer. The samples were prepared by UHV electron gun evaporation in the MBE laboratory of the Wigner Research Centre for Physics, Hungary. The nominal composition of the samples were MgO(100)/F/V(40 nm)/Cu(31nm)/Au(5nm). For the F layer Ni, Fe and Co were used respectively. In the following, samples will be noted as F$\mathrm{x}$, where F = Fe, Co, Ni and $\mathrm{x}$ is the thickness of the F layer in nanometers. The Cu layer was deposited onto the S/F bilayer to form a neutron waveguide structure with the MgO substrate to enhance the precision of the PNR measurements \cite{Khaydukov3}. The Au capping layer was used to prevent oxidation. The MgO (100) substrates were cleaned before deposition by rinsing in isopropanol in an ultrasonic bath and annealing at $600^\circ$C for 30 minutes in UHV. The maximum base pressure was less than $4.2\times10^{-10}$ mbar which did not increase above $6\times10^{-9}$ mbar during the evaporation. All layers were deposited at room temperature, except for the F layers which were deposited at $300^\circ$C. The deposition rates were 0.05, 0.2, 0.15 and 0.26 {\AA}/s for the F,  V, Cu and Au layers, respectively. Two samples were prepared simultaneously with the same nominal composition: one for PNR measurements (on $20\times 20\times 2$ mm$^2$ substrate) and one for SQUID and transport measurements (on $5\times 5\times 2$ mm$^2$ substrate).

\begin{figure*}[htb]
\centering
\includegraphics[width=2\columnwidth]{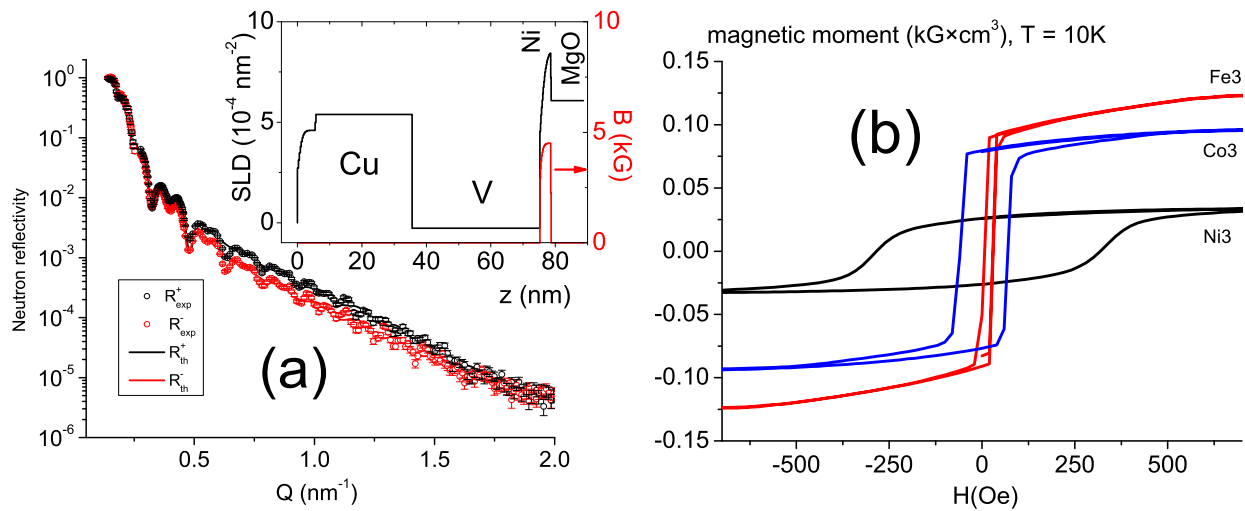}
\caption{
 (a) Experimental (dots) and calculated (lines) neutron-polarized reflectivity curves for the Ni3 sample measured at T = 5K and in H = 4.5kG. The corresponding model of the nuclear SLD and magnetic induction depth profiles is depicted in the inset. (b) Magnetic hysteresis loops measured at T = 10K on the same sample.
}
\label{Fig1}
\end{figure*}

The structural and magnetic characterization of the samples was performed by PNR. Some of the samples were measured on the GINA angle dispersive reflectometer (Budapest Neutron Center) \cite{Bottyan} at room temperature. These measurements were complemented later with low-temperature experiments at the NREX \cite{garching}, SuperAdam \cite{Devishvili} and V6 \cite{Paul} reflectometers. Additionally, the magnetic properties were measured using the SQUID magnetometer in the Max-Planck Institute for Solid State Research, Stuttgart, Germany. A typical reflectivity curve, measured at the NREX reflectometer on the Ni3 sample at $T = 5$K in a magnetic field of $H = 4.5$kOe is shown in Fig. \ref{Fig1}a. The curve is characterized by oscillations originating from the interference of neutron waves reflected from the different interfaces of the multilayer structure. The presence of a magnetic moment in the sample is indicated by the shift of the reflectivity of spin-up and spin-down polarized neutron beams ($R^+$ and $R^-$ curves in the figure).  The experimental curves were fitted to a layer model varying the thickness and the nuclear scattering length density (SLD) of the layers and the F layer magnetization. Parameters corresponding to the minimum $\chi^2$ were used to reconstruct the SLD and magnetization depth profiles (see inset in Fig. \ref{Fig1}a). According to the fit, the S and F layer thickness, $d_{S}$ and $d_{F}$, are within 10\% of the nominal values. The S/F interface is characterized by an rms roughness of $\delta_{SF} = 0.8$ nm. The nuclear SLD of the V and Ni layers were found to be close to the literature values. The magnetic profile of the structure is shown by red in the inset of Fig.\ref{Fig1}a. Above the $T_C$ of vanadium only the Ni layer showed a magnetic response with a magnetization of $M = 4.5$kG. This value is 92\% of the bulk nickel saturation magnetization and is in agreement with the saturation moment measured by SQUID at 10K (Fig. \ref{Fig1}b). The samples Fe3, Co3 and Fe1 were measured and analyzed in similar way (Table~\ref{Table 1}). As shown in Table~\ref{Table 1} the measured samples are characterized by S/F interfaces with $0 <  \delta_{SF} < 1$nm. The saturation magnetizations of the samples with $d_F = 3$nm are close to the bulk values (92\%, 98\% and 100\% for Ni3, Co3 and Fe3). The saturation magnetization of Fe1 is 20\% reduced as compared to the bulk, which may be explained by the reduced dimensionality.

\begin{figure*}[htb]
\centering
\includegraphics[width=2\columnwidth]{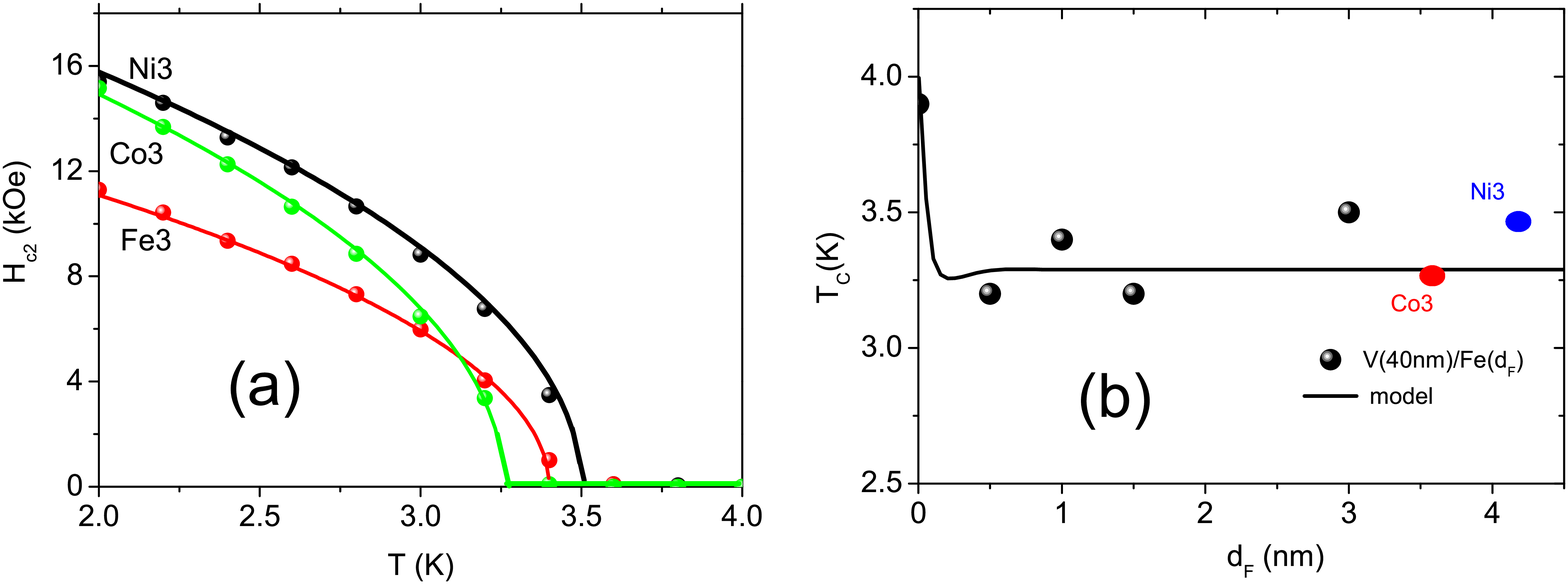}
\caption{
 (a) Temperature dependence of the upper critical field $H_{C2}$ of the different samples. the dots are experimental data, the lines are results of fits to Eq. (1). (b) Experimental (dots) $T_C(d_F)$ dependence. The line is the result of a model calculation described in the text.}
\label{Fig2}
\end{figure*}

The superconducting parameters of the S layers were determined using standard four-point DC electrical resistivity measurements with the magnetic field applied parallel to the sample surface. A series of temperature and magnetic field scans allowed us to determine the critical temperature $T_C$ and the upper critical field $H_{C2}$. The temperature dependence of the upper critical field $H_{C2}(T)$ of the 3 nm samples is shown in Fig.\ref{Fig2}a.

For a 2D thin film with the magnetic field applied parallel to the sample surface the temperature dependence of $H_{C2}(T)$ can be written as \cite{Tinkham},
\begin{equation}
H_{C2}(T) = \frac{\Phi_0}{2 \pi \xi_{GL}(0)}\frac{\sqrt{12}}{d_S}\sqrt{1-\frac{T}{T_C}},
\end{equation}
where $\Phi_0$ is the flux quantum, $\xi_{GL}$ is the Ginzburg-Landau correlation length related to Pippard's correlation length, $\xi_S$, by $\xi_S = 0.6 \xi_{GL}(0)$, and $T_C$ is the critical temperature in zero magnetic field. The fit of the curve $H_{C2}(T)$ using this expression and the values of $d_S$ measured by PNR allowed us to obtain the values of $T_C$ and $\xi_S$ for all samples. These values are presented in Table~\ref{Table 1}. Both the transition temperatures and the correlation lengths are close to the values previously reported in the literature \cite{Salikhov, Obi}. The magnetic field penetration depth $\lambda(0)$, presented in Table~\ref{Table 1}, was estimated from the expression $H_{C2}(0) \approx 5 H_\mathrm{bulk} \lambda(0)/d_S$ \cite{Schmidt}, where the critical field for bulk vanadium is $H_\mathrm{bulk} = 1.4$ kOe. From Table~\ref{Table 1} it can be concluded that the samples belong to the type II superconductor family with Ginzburg-Landau parameter $\kappa =  \lambda/\xi_S \sim 10$. Knowing $\xi_S$, the electron mean free path $l_S$ in the S layer may be estimated using the expression $\xi_S= \left(l_S \xi_\mathrm{BCS}/3.4 \right)^{1/2}$ where  $\xi_\mathrm{BCS} = 44$nm is the superconducting correlation length for the bulk vanadium. The resulting $l_S$ is of the order of 10 nm which indicates that the samples are closer to the diffusive limit ($l_S \ll \xi_\mathrm{BCS}$).


\begin{table*}
\caption{Structural, magnetic and superconducting parameters of the investigated samples. Here $d_F$ and $d_S$ are the thicknesses of the F and S layer, $M_\mathrm{sat}$ is the saturation magnetization at low temperatures, $\delta_{SF}$ is the rms roughness of the S/F interface, $T_C(0)$ and $H_{C2}(0)$ are the critical temperature and magnetic field, $\xi_S$ and $\lambda(0)$ are the superconducting correlation length and the magnetic penetration length. The entry "n/m" corresponds to a value not measured.}
\label{Table 1}
\begin{center}
\begin{tabular}{lcccccccr}
ID & $d_F$, nm & $M_\mathrm{sat}$, kG & $\delta_{SF}$, nm & $d_S$, nm & $T_C(0)$, K & $H_{C2}(0)$, kOe & $\xi_S$, nm & $\lambda(0)$, nm\\
Fe1 & 1.1 & 17.5 & 0.6 & 40.2 & 3.5 & n/m & n/m & n/m\\
Fe3 & 3.9 & 21.6 & 0.5 & 42.6 & 3.4 & 17.4 & 9.3 & 105.9\\
Co3 & 3.5 & 17.7 & 0.7 & 36.1 & 3.3 & 24.0 & 7.9 & 123.8\\
Ni3 & 4.1 & 5.7 & 0.8 & 41.2 & 3.5 & 24.0 & 6.9 & 141.3
\end{tabular}
\end{center}
\end{table*}


The $T_C(d_F)$ dependence in Fig. \ref{Fig2}b shows that the transition temperatures of the S/F bilayers are 10-20\% suppressed compared to $T_C = 4$K of the sample with similar $d_S = 40$ nm and $d_F = 0$. Such suppression is normally explained by the proximity effect, i.e. the leakage of Cooper pairs from a superconducting to a ferromagnetic layer. The experimental data were compared to a model based on the solution of the quasiclassical Usadel equation in the diffusive limit \cite{Fominov}. The solid curve depicted in Fig. \ref{Fig2}b was obtained for parameters $E_\mathrm{ex} = 800$ K, $\gamma = 1.2$ and $\gamma_B = 0.24$, where $E_\mathrm{ex}$ is the exchange field in the F layer, the dimensionless parameter $\gamma = \xi_S \sigma_F/ \xi_F \sigma_S$ determines the strength of the suppression of the superconductivity in the S layer near the interface compared to the bulk, and the dimensionless parameter $\gamma_B = R_T \sigma_F /\xi_F$ describes the effect of the interface barrier. Here $\sigma_{F,S}$ are the normal state conductivities of the F and S layers, $R_T$ is the interface resistance, and $\xi_{S,F}$ are the superconducting correlation lengths in the S, F layers, correspondingly. The $d_S$ and the $\xi_S$ values were taken from the PNR and the transport data, the value of the superconducting correlation length in the ferromagnet was assumed to be $\xi_F = 1$nm, which is a typical value for strong ferromagnets like iron, cobalt and nickel.

\begin{figure*}[htb]
\centering
\includegraphics[width=2\columnwidth]{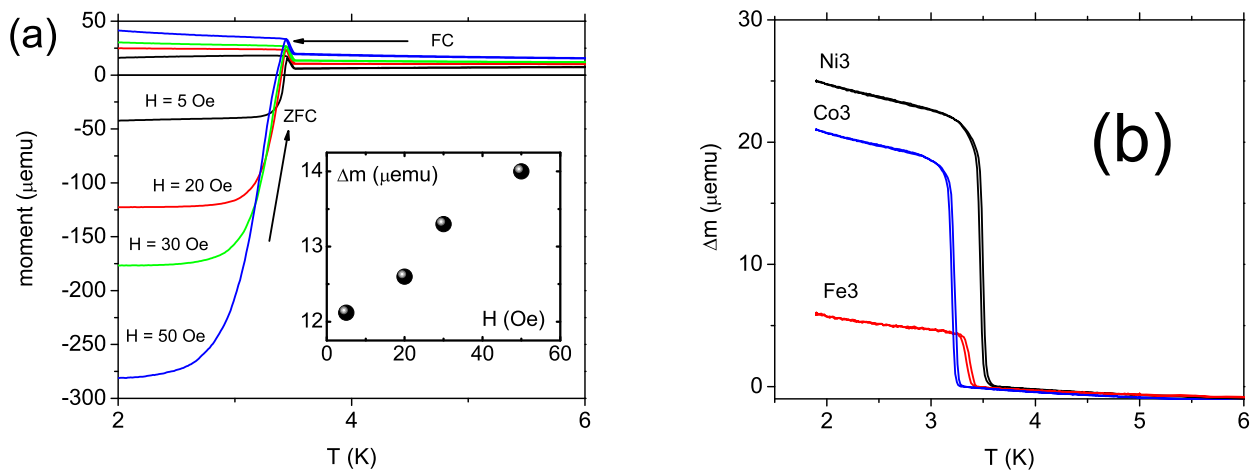}
\caption{
(a) Temperature dependence of the magnetic moment around $T_C$ measured on sample Fe1 in different magnetic fields and cooling regimes (FC = field cooled, ZFC = zero filed cooled). The inset shows the field dependence of the induced magnetic moment in the FC regime. (b) Temperature dependence of the induced magnetic moment for the samples with $d_F$ = 3nm in FC mode in a magnetic field of H = 10 Oe.
 }
\label{Fig3}
\end{figure*}

The temperature dependence of the magnetic moment in the vicinity of the superconducting transition was measured by SQUID magnetometry with the magnetic field applied in the sample plane. The SQUID measurements were performed as follows. At $T = 10$K the sample was saturated in a magnetic field of 1-2 kOe. In the following the field was released to zero and the sample was cooled down to $T = 2$K, where the working field was applied. The magnetic moment was measured while heating the sample to 10K (ZFC branch) and further cooling to 2K (FC branch). The temperature dependencies measured on the sample Fe1 at different magnetic fields are shown in Fig. \ref{Fig3}a. The ZFC curves are characterized by the decrease of the magnetic moment below $T_C$ due to the Meissner effect. The FC curves, on the other hand, show an increase of the magnetic moment. The value of the induced moment $\Delta m \equiv m(T<T_C)- m(T>T_C)$ in the FC regime depends linearly on the external field strength. Samples with $d_F = 3$nm show similar jump in the moments (Fig. \ref{Fig3}b). The magnitude of the jump is largest for Ni3, then Co3 and smallest for Fe3.

\begin{figure}[htb]
\centering
\includegraphics[width=\columnwidth]{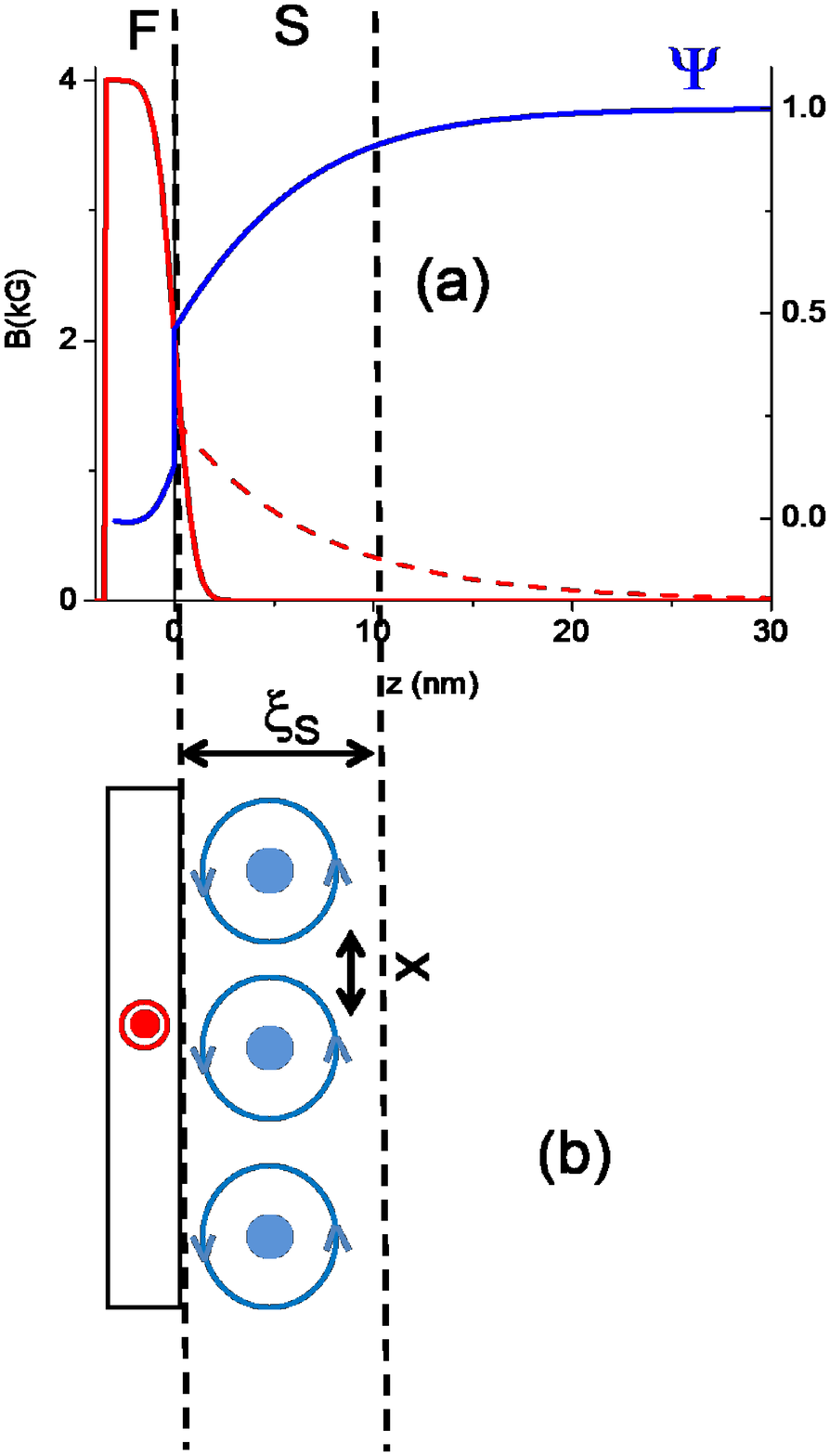}
\caption{
(Color online).
Depth profiles of the magnetization of the Ni3 sample above $T_C$ (solid red line), the proposed model of the induced magnetization below $T_C$ based on the previous results \cite{Khaydukov1,Khaydukov2} (dashed red line) and the density of superconducting pairs as it follows from the estimated parameters  $\gamma_B, \xi_S$ and  $\xi_F$. (solid blue line). The inset shows a sketch of the model. The blue circles represent a schematic row of vortices. The distance between the centers of the vortices is x, the direction of the currents flowing on the edge of the vortices are shown by the blue arrows.
 }
\label{Fig4}
\end{figure}

Thus, the SQUID data show that all of the measured samples exhibit an increase of the magnetic moment below $T_C$ after field cooling. A similar increase was observed by SQUID in our previous work on Cu(30nm)/V(40nm)/Fe(1nm) structure \cite{Khaydukov2}. There, by complementing the SQUID data with PNR measurements, a model was proposed where the increase of the magnetic moment was described as a magnetic sublayer in the S layer with a thickness of several nanometers and an induced magnetization of about 1 kG parallel to the magnetization of the F layer (Fig. \ref{Fig4}a). In the current study PNR measurements performed on big pieces also show similar features, i.e. change of the spin-flip scattering below $T_C$ similar to \cite{Khaydukov1} and shift of the spin asymmetry of non-spin-flip scattering similar to \cite{Khaydukov2}. However, the reproducibility of the PNR data, compared to the SQUID results was very poor. The magnetometry results were reproduced in several runs separated by a time period of a few months. The reason of this is not clear so far, but it may well be related to the difference in the area dimension. Future PNR experiments should be performed on the very same samples as were used for SQUID measurements.

To explain the increase of the magnetic moment of S/F bilayers upon passing below $T_C$ we propose the following model. Since vanadium is a type II superconductor, vortices may be present in the system. The size of a vortex core is comparable to $\xi_S$ and the magnetic induction of the core ($B_C$) is aligned parallel to the external field. The magnetic induction inside a single, isolated vortex core is $B_C \approx \Phi_0 \log (\kappa)/2 \pi \lambda^2$ \cite{Schmidt}. This estimation gives 1-2 kG for these systems, which is close to the induced magnetization observed previously by PNR. This allows the interpretation that the induced layer consists of a densely packed row of vortices in the vicinity of the S/F interface (Fig. \ref{Fig4}). During field cooling at temperatures just below $T_C$, vortices are created in the S layer. Upon further cooling well below $T_C$ only a single row of vortices directly adjacent to the S/F interface survives. Possible explanations of why and how the vortices are attracted and fixed to the S/F interface can be found in the literature. For example, a) attraction of the vortices to the bulk superconductor surface discovered by Bean and Livingston \cite{Bean}, b) pinning of the vortices to the S/F interface by roughness (which may be supported by the correlation of the induced moment and the roughness on the S/F interface, see Table~\ref{Table 1} and Fig. \ref{Fig3}b, c) general preference of the vortex location at the minimum of superconducting pair potential \cite{Lenk}, i.e. at the vicinity the S/F interface in our case (Fig. \ref{Fig4}a). Identifying the actual origin of vortex attraction is a subject of further studies.

Let us estimate the change of the magnetic moment $\Delta m$ in the S/F structure below $T_C$. The paramagnetic contribution, $\Delta m_V$, originates from the vortex cores, and the Meissner contribution, $\Delta m_m$, from the vortex-free part of the S layer. The paramagnetic contribution can be estimated as $\Delta m_V = B_C \xi_S C_V$, where $C_V = \xi_S/(\xi_S+x)$ is the linear concentration of the vortices at the S/F interface. Here $x$ is the distance between centers of neighboring vortices (Fig. \ref{Fig4}b).
The diamagnetic Meissner contribution can be calculated by integrating the depth profile of the Meissner response
\begin{equation}
B_M(z) = H \left ( \frac{\cosh (z/\lambda)}{\cosh (d_S/2 \lambda)} - 1 \right), \;\; z \in \left [ -\frac{d_S}{2}, \frac{d_S}{2}\right ].
\end{equation}

In our case $\lambda \gg d_S$ and the diamagnetic contribution is very small. For example, calculations using the experimentally obtained parameters for Ni3
and $H = 10$ Oe result in a paramagnetic response that is two orders of magnitude greater than the diamagnetic response. In order to describe the
experimentally observed $\Delta m$ a vortex concentration of about 80\% is required, although this model neglects the interactions between the
vortex cores. Furthermore, the increase of the induced moment with the external field (inset in Fig. \ref{Fig3}a) may be explained by the
increase of the vortex density with increasing field.

In conclusion, an increase of the magnetic moment in superconductor/ferromagnet (S/F) bilayers V(40nm)/F [F=Fe(1,3nm), Co(3nm), Ni(3nm)] was observed using SQUID magnetometry upon cooling below the superconducting transition temperature $T_C$ in magnetic fields of 10 Oe to 50 Oe applied parallel to the sample surface. To explain this effect a model based on a vortex row located at the S/F interface was proposed. According to the model the magnetic moment induced below $T_C$ consists of a paramagnetic contribution of the vortex cores and a diamagnetic contribution of the vortex-free region of the S layer. Since the thickness of the S layer was found to be 3-4 times less than the penetration depth of the magnetic field, this latter diamagnetic contribution is negligible. The model correctly accounts for the sign, the approximate magnitude and the field dependence of the paramagnetic and Meissner contributions of the induced magnetic moment upon passing the superconducting transition in superconductor/ ferromagnet bilayers.


\acknowledgments
The authors would like to thank A. A. Golubov, V. Zdravkov, R. Tidecks, G. Logvenov, H. U. Habermeier, and S. Soltan for fruitful discussion. This work is partially based upon experiments performed at the NREX instrument operated by Max-Planck Society at the Heinz Maier-Leibnitz Zentrum (MLZ), Garching, Germany. Support from DFG Collaborative Research Center TRR 80, RFBR (Projects 14-22-01007, 14-22-01063), BMBF (projects 05KN7PC1 and 05K10PC1) and Swedish Research Council is gratefully acknowledged. The results of the project T3-97 "Macroscopic Quantum Phenomena at Low Temperatures", carried out within the framework of the Basic Research Program at the National Research University Higher School of Economics (HSE) in 2016, are presented in this work.

\end{document}